
\magnification=\magstep1
\font\sixteenrm=cmr17 at 16pt
\def\bt{\beta}
\def\a{\alpha}          \def\w{\omega}         \def\sy{\psi}
\def\g{\gamma}          \def\W{\Omega}

\def\o{\theta}          \def\m{\mu}            
\def\O{\Theta}          \def\n{\nu}            \def\e{\epsilon}

\def\vf{\varphi}                         
\def\vo{\vartheta}                       
\def\ve{\varepsilon}

\def\^{\wedge}
\def\cd{\nabla}   

\def\DA{{\dot{A}}}
\def\DB{{\dot{B}}}

\def\DD{{\dot{D}}}
\footline={\hfil\tenrm\folio\hfil}
\line{\hfil{gr-qc/9403026}}
\bigskip
\line{\sixteenrm Some Spinor-Curvature Identities \hfil  }

\bigskip
\line{James M. Nester
\footnote{\dag}
{Supported by the
National Science Council of ROC under No.
 82-0208-M-008-013},
 Roh-Suan Tung${\dag}$
 and Vadim V. Zhytnikov
\footnote{\ddag}
{Supported by the
National Science Council of ROC under No.
 82-0208-M-008-070}
\footnote{*}
{Permanent address:
Nuclear Safety Institute,
Bolshaya Tulskaya 52,
Moscow 113191, Russia}
\hfil }   

\medskip
\line{Department of Physics, National Central University,
Chung-Li, Taiwan 32054  \hfil  } \bigskip\bigskip\noindent
\item{}{\bf Abstract:}
We describe a class of spinor-curvature identities
which exist for Riemannian or Riemann-Cartan geometries.
Each identity relates an expression quadratic
in the covariant derivative of a spinor field
with an expression linear in the curvature
plus an exact differential.
Certain special cases in 3 and 4 dimensions which have been
or could be used in applications to General Relativity are noted.

\item{}{ 02.40, 04.20  \hfil }


\bigskip\bigskip\noindent
Spinor techniques continue to provide new results in physics and
mathematics.
For any n-dimensional Riemann or Riemann-Cartan geometry with
a spinor structure we have found a certain class  of identities
each of which  relates an
expression quadratic in the covariant differential of a spinor field
with an expression linear in the curvature plus an exact differential.
We describe these identities in general and consider in particular
certain special cases in 3 and 4 dimensions which have application
to Einstein's gravity theory.

An orthonormal coframe field $\vo^a$ (note: the metric tensor
components $g_{ab}$ are constant) and
the metric compatible connection 1-form $\w^{ab}=-\w^{ba}$ are
``potentials'' for the {\it torsion} and {\it curvature} 2-forms:
$$\O^a:=d\vo^a+\w^a{}_b\^\vo^b,\qquad
\W^a{}_b:=d\w^a{}_b+\w^a{}_c\^\w^c{}_b.\eqno(1)$$
The dual n-k forms
$\eta^{ab\cdots}:=*(\vo^a\^\vo^b\cdots)$
are sometimes convenient.

A Clifford algebra
$\{1,\g_a,\cdots,\g_{ab\cdots c},\cdots,\g\}$,
 where
$\g_{ab\cdots c}:=\g_{[a}\g_b\cdots\g_{c]}$
and
$\g:=\g^1\g^2\cdots\g^n$, is
generated by ``Dirac matrices'' satisfying $\g_a\g_b+\g_b\g_a=2g_{ab}$.
Some of our results are most succinctly presented in terms of
the Clifford algebra valued forms:
 $$\eqalignno{
\vo:=\g_a\vo^a,\qquad&\w:={\textstyle{1\over4}}\g_{ab}\w^{ab} &(2)\cr
\O:=\g_a\O^a,\qquad&\W:={\textstyle{1\over4}}\g_{ab}\W^{ab},&(3)\cr}$$
In this notation eqs (1) become
$$\O:=D\vo:=d\vo+\w\^\vo+\vo\^\w,\qquad
\W:=d\w+\w\^\w.\eqno(4)$$
(A nice development of such ``Clifforms'' is
given by Dimakis and M\"uller-Hoissen 1991.)

The identities discussed here are all linear combinations of
expressions of the general type
 $$2D(\overline\sy A)\^ D(B\sy)\equiv d\{\overline\sy A\^ D(B\sy)
 -\ve D(\overline\sy A)\^B \sy\}
 -\ve 2\overline\sy A\^\W\^B\sy,
 \eqno(5)$$
where $\sy$ is a Dirac spinor,
$A$ and $B$ are matrix valued forms of rank $a,b$
and $\ve=(-1)^a$.
The general identity is easily established from
 $$d\{\overline\sy A\^ D(B\sy) -\ve D(\overline\sy A)\^B \sy\}
\equiv 2D(\overline\sy A)\^ D(B\sy)
+\ve\overline\sy A\^D^2(B\sy)-\ve D^2(\overline\sy A)\^B\sy,\eqno(6)$$
using $D^2(B\sy)=\W B\sy$ and $D^2(\overline\sy A)=-\overline\sy A\W$.
The simplest case, $A=B=1$, yields an identity with the full
curvature 2-form:
$$2D\overline\sy\^D\sy\equiv d(\overline\sy D\sy-D\overline\sy \sy)
-{\textstyle{1\over2}}\W^{ab}\overline\sy\g_{ab}\sy.\eqno(7)$$

It should be noted that identities of this general form considered
here are not
confined to spinor fields, for example for 2 vector fields we have
$$\eqalign{
DW_\a\^DV^\a&\equiv
d(W_\a DV^\a-V^\a DW_\a)-(W_\a D^2V^\a-V^\a D^2W_\a) \cr
&\equiv d(W_\a DV^\a-V^\a DW_\a)-2W_\a\W^\a{}_\bt V^\bt.\cr}\eqno(8)$$
Here, however, we consider spinor fields and focus on the cases
where $A$ and $B$ in eq (4) are simple combinations naturally
constructed from $\vo$, their matrix-wedge
products $\vo^K:=\vo\^\vo\^\dots$
($K$ factors) and the Hodge dual $*$ because,
for suitable
choices of such $A$ and $B$, we can project out curvature
combinations such as
the  Einstein tensor or the scalar curvature; we do not
know how to get these using vectors or tensors.

The 3 and 4 dimensional versions of such identities have
application to gravity theory.
In particular they can be used to replace the curvature terms in
Lagrangians and Hamiltonians.  This can be advantageous because the
quadratic spinor expressions have faster asymptotic fall off.

The first construction of this type contained the 4-dimensional
Einstein tensor and was used in the Witten positive
energy proof (Witten 1981, Nester 1981).
Subsequently it was shown that the {\it Hamiltonian density}
for Einstein gravity
could be expressed as a 4-covariant quadratic spinor 3-form
(Nester 1984):
$$
{\cal H}(\sy)
:= 2\{D(\overline\sy\g_5\vo)\^D\sy-D\overline\sy\^D(\g_5\vo\sy)\}
       \equiv 2 N^\mu G^\nu_{\,\mu} \eta_\nu +dB, \eqno(9)$$
where
$$B:=\overline\sy\g_5\vo\^D\sy+D(\overline\sy\g_5\vo)\sy
  -\overline\sy D(\g_5\vo\sy)+D\overline\sy\^(\g_5\vo\sy),\eqno(10)$$
and
$N^\m=\overline\sy\g^\m\sy$.  The key is that
$$\W^{\m\n}\^\overline\sy\g_5(\vo\g_{\m\n}+\g_{\m\n}\vo)\sy\equiv
  \W^{\m\n}\^\eta_{\m\n\a}\overline\sy\g^\a\sy\equiv
  G^\n{}_\m\eta_\n\overline\sy\g^\m\sy.\eqno(11)$$
The Hamiltonian density (9) can be decomposed, with respect to the
normal to any
spacelike hypersurface, into positive and negative definite parts
and is locally non-negative
if
$\sy$
satisfies  the
Witten  equation
$    \g^a D_a \sy=0, $
 thereby permitting a positive energy proof and more: a
non-negative
``localization'' of gravitational energy.

More generally for an n-dimensional Riemanian space the Einstein
tensor (for $n>3$) appears in the ``Hamiltonian'' $n-1$ form
(Mason and Frauendiener 1990):
 $$\eqalignno{
{\cal H}(\sy)
         &:=  4D\overline\sy\^\g\vo^{(n-3)}\^D\sy \cr
         &\equiv 2 \overline\sy\g^b\sy G^a{}_b \eta_a +
d\{\overline\sy\g\vo^{(n-3)}\^D\sy
+D\overline\sy\^\g\vo^{(n-3)}\sy\}.&(12)\cr}$$
The corresponding Riemann-Cartan expressions are considerably more
complicated as it is necessary to take into account non-vanishing
contributions
from $D\vo=\O$ on the left hand side and grade 5 Clifford algebra terms
proportional to
$D\O^a\equiv\W^a{}_b\^\vo^b$ on the right hand side;
the general pattern is
binomial, in particular $\overline\sy\g(\W\vo\vo+2\vo\W\vo+\vo\vo\W)\sy$
projects out the Einstein 4-form in 5 dimensions; for 6 dimensions
$\overline\sy\g(\W\vo\vo\vo+3\vo\W\vo\vo+3\vo\vo\W\vo+\W\vo\vo\vo)\sy$
gives the Einstein 5-form, etc.
By the way, the  rank of such  identities is not as restricted as our
discussion has
indicated for we could take $\sy$ to be a spinor valued differential
form of some suitable rank.

The next application came up in
a new gravitational energy positivity proof
and localization using 3-dimensional spinors (Nester and Tung 1993).
The {\it scalar curvature} term
and the boundary term in the ADM Hamiltonian were replaced using
the new {\it spinor identity}
$$
 2[\cd( \vf^{\dag}i\vo) \^\cd\vf -\cd \vf^{\dag}\^\cd(i\vo\vf)]
\equiv  d B -  (\vf^{\dag}\vf) \W^{ij} \^ \eta_{ij}, \eqno(13)$$
where
$$B:=  \vf^{\dag}i\vo\^\cd\vf -\vf^{\dag}\cd(i\vo\vf)
       +\cd(\vf^{\dag}i\vo)\vf +(\cd\vf^{\dag})\^i\vo\vf. \eqno(14)$$
The key detail is
$$\vf^{\dag}\vf R\sqrt{g}d^3x
\equiv\vf^{\dag}\vf\W^{ij}\^\eta_{ij}
\equiv \W^{ij}\^\vf^{\dag}i(\vo\g_{ij}+\g_{ij}\vo)\vf.\eqno(15)$$

Generalizing, we find an identity which contains the Riemannian scalar
curvature n-form: $R\eta=\W^{ab}\^\eta_{ab}$,
$$2D\overline\sy \^\g\vo^{(n-2)}\^D\sy\equiv
 d\{\overline\sy\g\vo^{(n-2)}\^D\sy-D\overline\sy\^\g\vo^{(n-2)}\sy\}
-\overline\sy\sy\W^{ab}\^\eta_{ab}.\eqno(16)$$
The corresponding Reimann-Cartan identity is more complicated but
again follows a binomial pattern, the general structure can be
inferred from
the special cases in low dimensions.
For 3 dimensions we have eq (13), for 5
dimensions the Clifform combination which projects out purely
the scalar curvature 5-form is again $\overline\sy\g(\W\vo\vo\vo
+3\vo\W\o\vo+3\vo\vo\W\vo+\vo\vo\vo\W)\sy$, for 4 dimensions
it is $\overline\sy\g(\W\vo\vo+2\vo\W\vo+\vo\vo\W)\sy$
so that we have
in detail
$$\eqalignno{
D(\overline\sy\g\vo^2)& \^D\sy
-2D(\overline\sy\g\vo)\^D(\vo\sy)
+D\overline\sy\g\^D(\vo^2\sy)
\equiv
-R\overline\sy\sy\eta\cr
&+d\{-D\overline\sy\^\g\vo^2\sy-D(\overline\sy\g\vo)\^\vo\sy
-\overline\sy\g\vo\^D(\vo\sy) +\overline\sy\g\vo^2\^D\sy\},&(17)\cr}$$
for the 4-dimensional identity.
This expression illustrates how much more complicated the identities
are with torsion.  The Riemannian version is simply
$$4D\overline\sy\^\gamma\vo^2\^D\sy\equiv
-R\overline\sy\sy\eta+
2d\{-D\overline\sy\^\gamma\vo^2\sy +\overline\sy\gamma\vo^2\^D\sy\}.
\eqno(18)$$
  These 4-dimensional scalar curvature identities invite
applications where the Einstein-Hilbert La\-grang\-ian
 is replaced by a quad\-ra\-tic spinor Lagrangian 4-form.

The higher dimensional identities could also be useful for gravity
applications,
e.g., unified field theories of the generalized Kaluza-Klein type.
Moreover, a 5-dimensional geometry is useful for embedding
4 geometries with
a cosmological constant.  However, most gravity applications are
in 3 and 4 dimensions.

In 3-dimensions, up to multiples of $i$ there are
only 2 interesting cases
$D\vf^{\dag}\^D\vf$,  which gives the full curvature 2-form (7), and
the combination (13) which gives the scalar curvature.

The basic 4-dimensional cases are as follows.
First the
2-form $D\overline\sy\^D\sy$ is related to the full curvature tensor
according to (7).
Next we have the Einstein tensor (2).  And the scalar curvature (17).
Moreover we can insert an additional $\g_5$ into all of these.

We have found the Dirac spinor notation most suitable for the general
case. However, for the important special case of 4 dimensions (and 3
dimensions) another spinor
notation is also popular (Penrose and Rindler 1986).
The basic identities in this notation have the form
(Riemannian geometry):
$$
2D\vf_A\^D\chi^A\equiv
d(\vf_AD\chi^A-D\vf_A\chi^A)
+2\vf^A\chi^B\W_{AB},
\eqno(19)$$
$$
2D\vf_A\^\o^{A\DB}\^D\chi_{\DB}\equiv d[
\vf_A\o^{A\DB}\^D\chi_{\DB}
-D\vf_A\^\o^{A\DB}\chi_{\DB}]
- 2\vf_A\chi_\DB\W^\DB_\DD\^\o^{A\DD},
\eqno(20)$$
$$
\eqalignno{
2D\vf^A\^S_{AB}\^D\chi^B &\equiv
d[\vf^AS_{AB}\^D\chi^B
-D\vf^A\^S_{AB}\chi^B]
-2\vf^A\chi^BS_{AM}\^\W^M_B  & \cr
&\equiv
d[\vf^AS_{AB}\^D\chi^B
-D\vf^A\^S_{AB}\chi^B]
-{i\over4}R\vf^A\chi_A\eta,
&(21)\cr}$$
where $\vf^A$ and $\chi^A$ are any two-component Weyl spinors,
$S_{AB}$ is the self dual spinorial 2-form
$S_{AB}:= {1\over2}\o_A{}^\DD\^\o_{B\DD}$ and
$\W_{AB}$ is the spinorial curvature.

Four dimensional Dirac spinors and matrices can be expressed via
Weyl spinors with the help of the relations:
$$
\overline\sy=(\chi_A\quad\vf^\DA),\qquad
\sy={\vf^A\choose\chi_\DA},\qquad
\gamma^a=\sqrt2\left(
\matrix{0&\sigma^{aA\DB}\cr
\sigma^a_{B\DA}&0 }
\right),
\eqno(22)$$
where $\sigma^a{}_{A\DB}$ satisfies the identity
$\sigma^a{}_{A\DB}\sigma_{aC\DD}=\ve_{AC}\ve_{\DB\DD}$.

Identities (19)--(21) are complex, but with the help of (22)
one can easily conclude that eq (21) is the imaginary part of
the (18) and the analog of (18) without $\gamma$ is the
real part of (21). Similarly, eq (7) is the real part of
(19).

We  also note that
such identities are not restricted only to the Riemann or
Riemann-Cartan connection.
In 4 dimensions we may, for example,
replace $\w^{\a\bt}$ and $\W^{\a\bt}$
with their self dual combinations:
$A^{\a\bt}:={1\over2}(\w^{\a\bt}
+{1\over2}i\e^{\a\bt}{}_{\m\n}\w^{\m\n})$ and
$F^{\a\bt}:={1\over2}(\W^{\a\bt}
+{1\over2}i\e^{\a\bt}{}_{\m\n}\W^{\m\n})$
(for an application see Nester, Tung and Zhang 1993).
Moreover, similar identities apply to Yang-Mills connections,
for example
$$2D\overline\sy\^D\sy\equiv
 d(\overline\sy D\sy-(D\overline\sy)\sy)-2F^p\overline\sy T_p\sy,
\eqno(23)$$
where $D\sy:=d\sy+A^pT_p\sy$ and $F^p:=dA^p+\e^p{}_{qr}A^q\^A^r$ is the
Yang-Mills field strength.

We have focused on identities linear in curvature. However,
identities of the type discussed here need not be so restricted;
we may choose, for example, $A=\W^{ab}\g_{ab}$ in eq (5) to obtain
an identity involving quadratic curvature terms.
Clearly there are many other
possibilities.  We are not yet aware of any applications for such
higher order identities.

\vfil
\bigskip\noindent 
{\bf References}
\medskip 
\parindent = 0pt
\def\Fullout{
 \rightskip=0pt
 \spaceskip=0pt
 \xspaceskip=0pt }
\def\bb{
   \Fullout
  \hangindent=1pc
  \hangafter=1 }
\bb  Dimakis A and M\"uller-Hoissen 1991
         {\it Class. Quant. Grav.} {\bf 8} 2093

\bb  Mason L J and Frauendiener J 1990
         in {\it Twistors in Mathematics and Physics}
         ed R Baston and T Bailey
         (Cambridge : Cambridge University Press) pp 189

\bb   Nester J M 1981
         {\it Phys. Lett. } {\bf 83A} 241

\bb   Nester J M 1984
         in {\it Asymptotic Behavior of Mass and Space-time Geometry},
         (Lecture Notes in Physics {\bf 202}) ed F Flaherty
         (Berlin : Springer) pp 155

\bb  Nester J M and Tung R S 1994
       Another positivity proof and gravitational energy localizations.
       (gr-qc/9401002) to appear in {\it Phys. Rev. D. }

\bb   Nester J M, Tung R S and Zhang Y Z 1994
       Ashtekar's new variables and positive energy. (gr-qc/9401004)
      to appear in {\it Class. Quant. Grav.}

\bb   Penrose R and Rindler W 1986
 {\it Spinors and Space-Time}
 (Cambridge : Cambridge University Press)

\bb   Witten E 1981
         {\it Comm. Math. Phys. } {\bf 80} 381

\bye